\documentclass[a4paper]{torque2026}

\usepackage{graphicx}
\usepackage{amsmath}
\usepackage{float}
\usepackage{natbib}
\usepackage[hyphens]{url}
\usepackage{hyperref}

\begin{document}

\setlength{\footskip}{20pt}
\pagestyle{plain}

\title{Evaluating the Role of Blockage Deficit Models in Robust Wind Farm Design}

\author{Manh Cuong Ngo$^{1}$ and Alexander Meyer Forsting$^{2}$}

\address{$^1$Department of Ecoscience, Aarhus University, Aarhus, Denmark}
\address{$^2$Department of Wind and Energy Systems, Technical University of Denmark, Roskilde, Denmark}

\ead{nm.cuong@ecos.au.dk}

\vspace{10pt}

\begingroup
\small
\noindent This is the Accepted Manuscript version of an article accepted for publication in \emph{IOP Science Journal of Physics: Conference Series 3224 032072}. IOP Publishing Ltd is not responsible for any errors or omissions in this version of the manuscript or any version derived from it. This Accepted Manuscript is published under a Creative Commons Attribution 4.0 International licence (CC BY 4.0). The Version of Record is available online at \href{https://doi.org/10.1088/1742-6596/3224/3/032072}{https://doi.org/10.1088/1742-6596/3224/3/032072}.

\vspace{6pt}

\noindent This arXiv version has been reformatted in \LaTeX\ and includes minor editorial
and structural revisions. The reported results and conclusions are unchanged from
the Version of Record.
\par
\endgroup

\begin{abstract}
 Uncertainty in wind farm layout optimization regarding model choice and the impact of global blockage always exists. This paper evaluates these interactions by extending a multi-objective approach that maximizes mean Annual Energy Production (AEP) from a model ensemble while minimizing their variance. By incorporating the Self Similar blockage model into an ensemble of five wake models, we assess the impact of blockage physics on layout robustness. Results from a linear mixed-effects model indicate that including blockage leads to a non-significant average AEP reduction of 0.0624 GWh ($p$-value = 0.323). Conversely, it caused a highly statistically significant increase in uncertainty, with model disagreement rising by 0.177 GWh ($p$-value < 0.001). Additionally, computational runtime increased nearly nine times. These findings highlight an accuracy vs. certainty paradox, where theoretically necessary physics can compromise model consensus if implemented without re-calibration. Ultimately, this work suggests that simple blockage couplings act as a diagnostic for model incompatibility, emphasizing the necessity of careful model tuning for wind farm design. 
\end{abstract}

\section{Introduction}
In Europe, the EU set a goal of attaining 425GW of wind energy in 2030. When planning wind farms, optimizing wind turbines position for economic values and energy efficiency is a critical mission, while also highly complex. It is due to aerodynamic losses, specifically wake effects and global blockage, which penalize high-density turbine spacing. Given the massive capital required for offshore deployment, possessing good models for estimating uncertainty would be essential to make suitable decisions.

Due to the complexity of aerodynamic phenomena within wind farms, there exists a variety of wake models with different levels of fidelity \cite{Mortensen2015}. Higher-fidelity models require more demanding computation, leading to a practical reliance on simpler engineering models. This leads to uncertainty when selecting the suitable wake effect models, in which different models can lead to different optimal locations of wind turbines. The problem has not been focused on until recently, when O’Neil et al. (2025) \cite{Oneill2025} proposed a method of multi-objective optimization by maximizing Annual Energy Production (AEP) of five popular wake effect models while minimizing the AEP variance.

Beyond wake effects, predicting farm performance requires accounting for upstream flow deceleration. In this study, we focus on the accumulation of individual turbine induction zones, often referred to as wind farm blockage. The global blockage effects, which are driven by atmospheric thermal stratification and boundary layer interactions \cite{Gribben2019, Stipa2023, Devesse2024a, Devesse2024b}, are beyond the scope of our work.

Recent studies \cite{Stipa2023, Devesse2024a, Devesse2024b, Branlard2020} and industry initiatives, such as the Global Blockage Effect in Offshore Wind (GloBE) campaign \cite{RWE2023}, have emphasized that blockage models should not be treated as independent add-ons. They state that wake models must be recalibrated when run iteratively with blockage models to ensure physical consistency. However, in many academic and early-stage development contexts, high-fidelity SCADA data required for such simultaneous calibration is unavailable. This often leads to a simple inconsistent coupling, where blockage models are applied to engineering wake models that were originally calibrated for wake-only scenarios.

While it is understood qualitatively that this can introduce errors, there is a lack of rigorous statistical methods to quantify the specific impact of this coupling on model consensus. In this study, we address this gap by extending the multi-model optimization method in O’Neil et al. \cite{Oneill2025} by including the Self Similar blockage deficit model \cite{Branlard2015, Forsting2015} to investigate its impact on layout performance and robustness using hierarchical models \cite{Fisher1919}. We use this framework to quantify exactly how much additional uncertainty, measured as the disagreement between wake models, is introduced when blockage physics are added without the requisite re-calibration.

\section{Method}

\subsection{Optimization and Data Generation}

To investigate the impact of blockage models, we formulated our question as an optimization problem of maximizing the Annual Energy Production (AEP) $\mu$ while minimizing the variance of AEP $\sigma^2$ between them. The objective function is:
\begin{equation}
	\begin{split}
		\text{Minimize} \quad & F(\mu(x,y), \sigma(x,y)) := -w \cdot \mu(x,y) + (1 - w) \cdot \sigma^2(x,y) \\
		\text{subject to} \quad & (x_i - x_j)^2 + (y_i - y_j)^2 \geq (N \cdot D)^2 \quad \forall i \neq j
	\end{split}
\end{equation}
where $x, y$ are the coordinates of the wind turbine, $D$ is the rotor diameter, and $N$ is a multiplying constant. The weighting parameter $w$ controls the trade-off between maximizing production and ensuring model agreement (robustness).

Specially in our case, following the syntax of O’Neil et al. \cite{Oneill2025}, we defined $\mu(x,y)$ as the mean of AEPs of five different wake deficit models TurbOPark Gaussian \cite{Nygaard2022}, Zong Gaussian \cite{Zong2020}, Blondel Gaussian \cite{Blondel2020}, Bastankhah Gaussian Deficit \cite{Bastankhah2014}, and Jensen \cite{Jensen1983}
\begin{equation}
	\mu(x,y) = \frac{1}{5} \sum_{i=1}^5 AEP_i
\end{equation}
and $\sigma^2 (x,y)$ is the variance of AEPs accross the five models above.

The ensemble of wake models used in this study is not considered the absolute ground truth, which can only be established via high-fidelity SCADA data or Computational Fluid Dynamics. Instead, the disagreement between the models represents systematic uncertainty, the structural uncertainty inherent in relying on simplified engineering approximations. While models such as the Gaussian and NOJ formulations are industry standards with a high degree of confidence for baseline AEP estimation, their divergence when subjected to modified inflow (via the blockage model) serves as a quantitative proxy for the underlying physical uncertainty in the coupled system.

\noindent {\bf The Role of the Weighting Parameter ($w$)}

In Equation (1), the scalar $w$ ($0 \leq w \leq 1$) is the critical hyperparameter that defines the optimization goal. When $w \rightarrow 1$, the optimizer ignores uncertainty and focuses solely on maximizing AEP. When $w \rightarrow 0$, the optimizer focuses purely on minimizing variance (robustness).

\vspace*{6pt}
\noindent {\bf Sampling Strategy}

To map the Pareto front efficiently, we performed the optimization for 24 distinct values of $w$ to balance between statistical sufficiency for the model and the computation efforts. Instead of linear spacing, we used a Chebyshev cosine mapping to cluster values near $w = 0$ and $w = 1$, capturing the knee of the Pareto curve where trade-offs are significant.
\begin{equation}
	w_i = \frac{1}{2} \left( 1 - \cos \frac{i\pi}{23} \right) \quad \text{for} \quad i = 0, \dots, 23
\end{equation}

In multi-objective optimization, small changes in weight can sometimes lead to large jumps in the final design. In constrast, large changes in weight might lead to no change at all \cite{Das1997, Marler2010}. Thus, in this study, we explore a non-uniform sampling method, which focuses on the endpoints of the weighting range ($w \approx 0$ and $w \approx 1$). It could allows us to trace the sharp curve (the knee) of the optimal solution set more precisely than a uniform grid would allow (Figure 1).

\begin{figure}[H]
	\centering
	\includegraphics[width=0.8\textwidth]{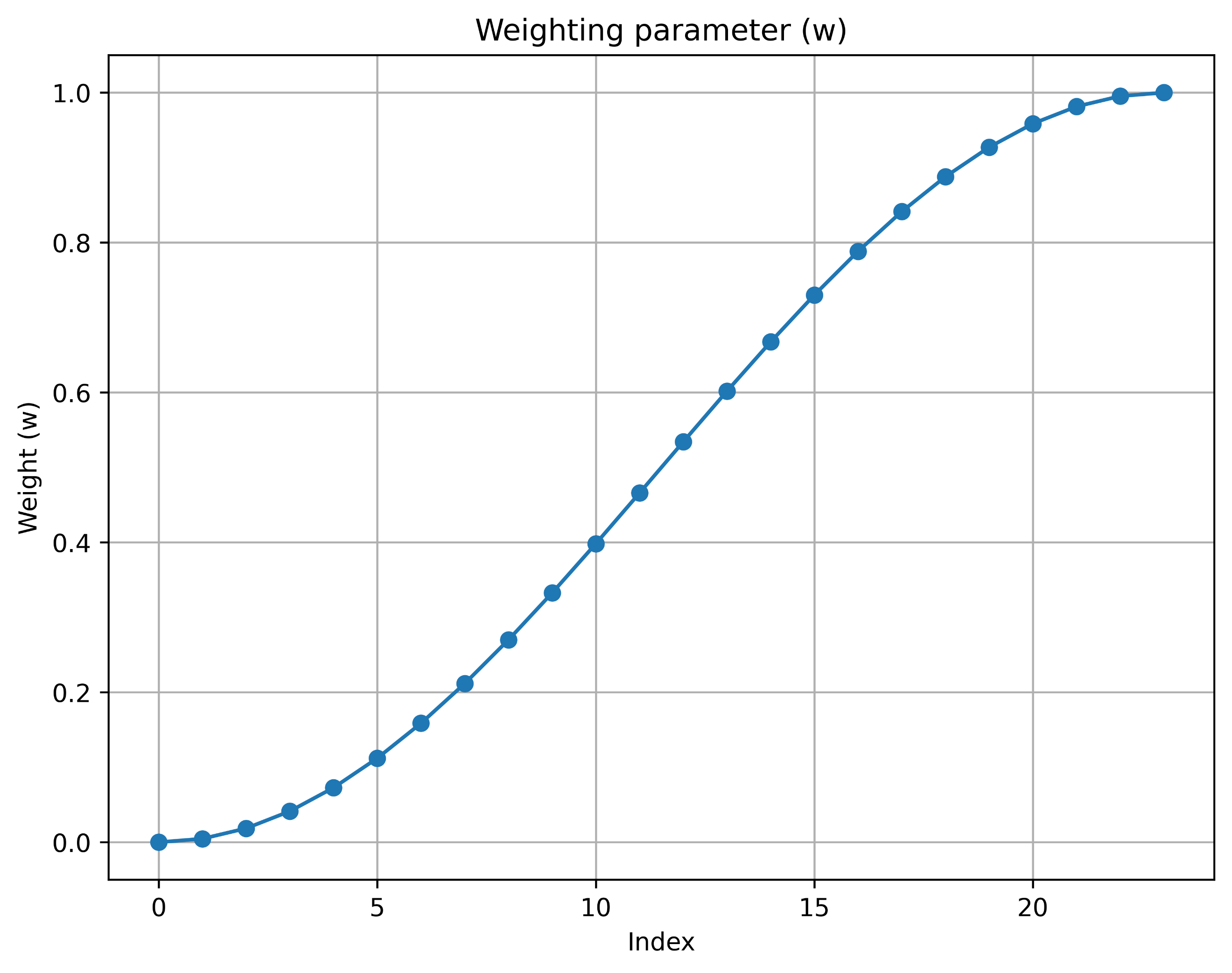}
	\caption{Weight parameters $w$ following Chebyshev cosine mapping.}
\end{figure}

\vspace*{6pt}
\noindent {\bf Case Study Parameters}

We chose 10 Vestas V80-2.0 MW turbines with diameter $D = 80$ m, and Hub Height $H = 70$ m. The wind climate was based on the Horns Rev 1 site with a mean wind speed of 9.02 m/s and turbulence intensity of 10\% (Weibull distribution with scale parameter $A \approx 10.16$ m/s and shape parameter $k \approx 2$) (Figure 2). Since Horns Rev 1 site is designed to 80 turbines, we restricted our area of interest to the site's center, covering 12.5\% (1/8) of the original area \cite{Hansen2012}. This ensures that wake and blockage interactions remain physically relevant.

\begin{figure}[H]
	\centering
	\includegraphics[width=0.95\textwidth]{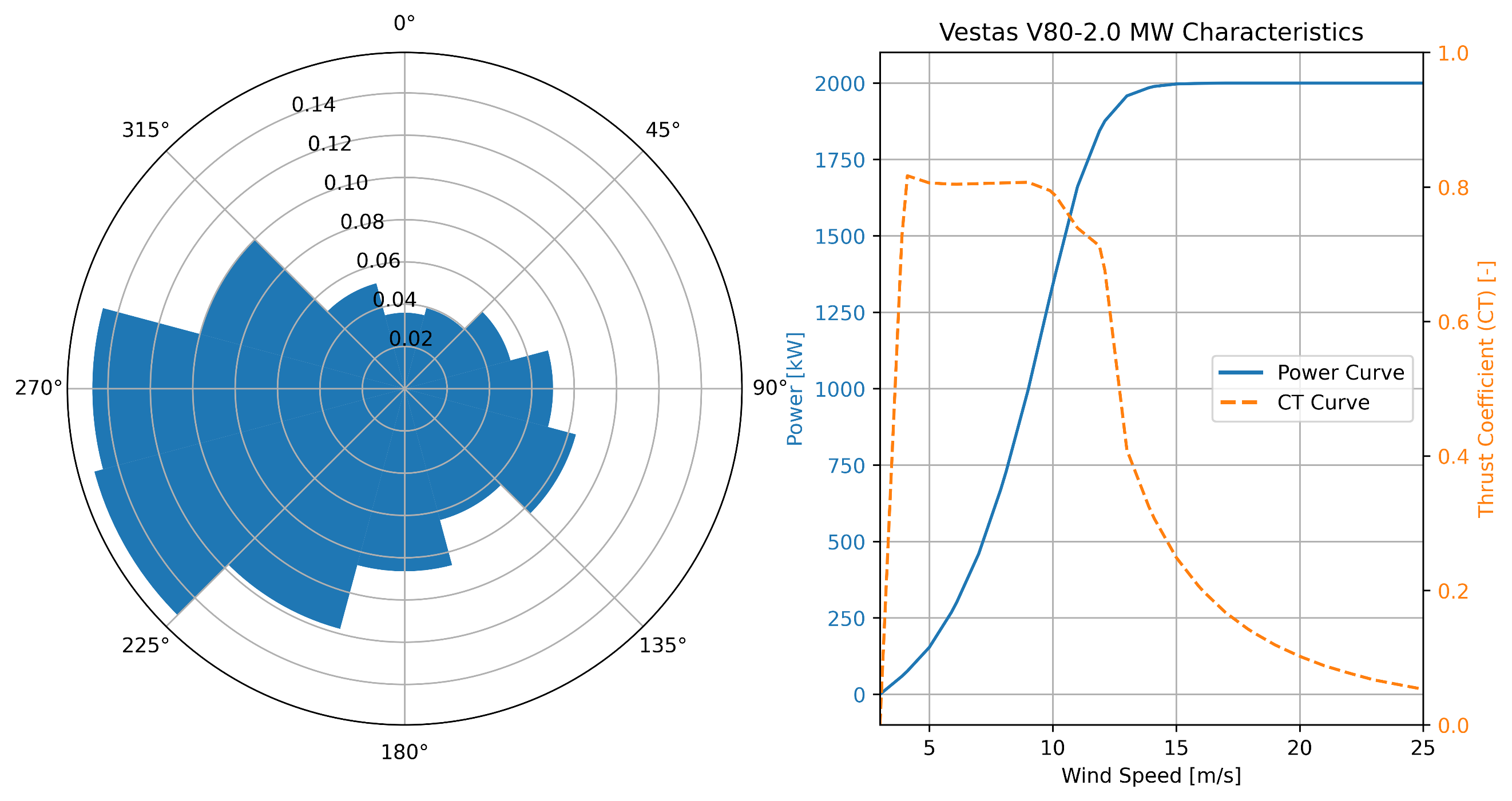}
	\caption{a) Wind Direction and b) Power \& CT Curve.}
\end{figure}

\vspace*{6pt}
\noindent {\bf Optimization Procedure}

We performed the optimization process with the SLSQP gradient-based optimizer. TopFarm’s SmartStart algorithm was used to ensure a valid initial state. For the flow field calculations, we selected the PropagateUpDownIterative solver. In standard engineering models, wake deficits are evaluated in a strictly downstream sequence. However, physically capturing blockage directs that downstream turbines also decelerate the inflow of turbines positioned upstream. This mutual two-way aerodynamic coupling cannot be solved in a single pass and requires an iterative solution. The PropagateUpDownIterative solver attains this by alternating downstream wake propagation with upstream blockage propagation until the local wind speeds converge. As part of this strategy, we restricted the blockage model to exclusively upstream interactions to prevent the unphysical double-counting of downstream energy losses already captured by the wake models. This solver ultimately balances the strict physical requirement of modeling upstream feedback with the computational efficiency demanded by an iterative gradient-based optimizer.

Turbine locations were limited by the site boundaries and a minimum spacing of two times the rotor diameter. Due to the limit of computing resources, we limited the study to 10 turbines while still sufficiently demonstrating the uncertainty reduction methodology.

\vspace*{6pt}
\noindent {\bf Simulation Setup}

We implemented the optimization process in Python with Topfarm \cite{Riva2023} and PyWake \cite{Pedersen2023}. We discretized the wind resource into 360 wind direction sector ($1 \deg$ resolution) and 23 wind speed bins ($1$ m/s resolution) with a uniform wind shear profile (no vertical shear). We selected five distinct wake deficit models with the appropriate superposition models: Linear Sum for Zong and Blondel models, and Squared Sum for TurboOPark \cite{Nygaard2022}, Bastankhah Gaussian Deflicit \cite{Bastankhah2014}, and Jensen models \cite{Jensen1983}. The suitable rotor-averaging methods were chosen separately for each of them: area overlap for Jensen model, Gaussian overlap for TurboPark, and center-line sampling for the rest (Zong, Bondel, and Bastankhah).

\subsection{Statistical framework}

This study extends the multi-model optimization framework of O'Neil et al. \cite{Oneill2025} by incorporating blockage models to investigate their impact on layout performance and robustness. Among various different blockage deficit models, here we chose the standard Self Similar model \cite{Forsting2015}. Physically, this model uses an analytical formulation derived from curve fits to high-fidelity Computational Fluid Dynamics to describe both the axial and radial deceleration of the wind approaching a rotor. We specifically chose this model over vortex-cylinder alternatives because its purely analytical, integral-free formulation provides the computational efficiency for our optimization framework. To rigorously quantify the impact of blockage physics, we should move beyond visual comparisons of Pareto fronts, for example with a statistical method that accounts for the structure of our data.

\vspace*{6pt}
{\bf Motivation for Hierarchical Modeling}

Our dataset consists of paired observations. A layout optimized with $w = 0.1$ is structurally different from one optimized with $w = 0.9$ (the latter prioritizes minimizing wake losses, resulting in a distinct spatial configuration). Comparing the Blockage result at $w = 0.9$ directly to the No Blockage result at $w = 0.1$ would be invalid because the underlying turbine positions are not the same. Furthermore, simply conducting 24 separate t-tests (one for each $w$) would lead to statistical errors (multiple comparisons problem) or fail to capture the global trend.

To address this, we employed a Linear Mixed-Effects Model (LMM), also known as a hierarchical model \cite{Fisher1919}. This approach allows us to analyze all 24 data points simultaneously while respecting the paired nature of the experiment.

\vspace*{6pt}
{\bf Model Specification}

We treat the optimization weight w not as a number, but as a grouping factor (Random Effect). The inclusion of blockage physics is treated as a fixed effect (the treatment). The model is specified as

\begin{equation}
	\text{AEP}_i = \beta_0 + \beta_1 \times \text{ModelType}_i + u_{w_i} + \epsilon_i
\end{equation}
where
\begin{itemize}
	\item $i$ is the individual observation index ($i = 1, \dots ,48$), representing the results across the 24 objective weights and two model configurations.
	\item $\beta_0$ is the global intercept (baseline AEP) (unitless).
	\item $\beta_1$ is the Fixed Effect of interest. It represents the average shift in AEP caused by including the blockage model, holding the optimization weight constant (unitless).
	\item $u_{w_i}$ is the Random Intercept for group w. This term captures the fact that AEP naturally varies depending on where we are on the Pareto front (high w yields high AEP, low w yields low AEP).
	\item $\epsilon_i$ is the residual error.
\end{itemize}

In this framework, the objective function relies on the mean AEP of the ensemble. We acknowledge that the mean is sensitive to extreme values: if a single wake model reacts unstably to the blockage coupling, it can disproportionately shift the mean $\mu(x,y)$ and the fixed effect estimate $\beta_1$. Applying a median shift instead of a mean shift could theoretically provide a more robust central tendency that ignores outlier models. However, the mean was intentionally selected to ensure a smooth, continuously differentiable objective function, which is a mathematical requirement for the gradient-based SLSQP optimizer used in this framework.

Here, $w$ serves as the grouping variable for the random effect, accounting for the paired nature of the data (comparing the same point on the trade-off curve with and without blockage). The model can be formulated using the syntax of the package \texttt{lme4} \cite{Bates2015} in R \cite{Rcoreteam2022}.
\begin{equation}
	\text{AEP} \sim \text{ModelType} + (1 | w)
\end{equation}

In the formula above, AEP ~ ModelType represents the fixed effect (Blockage vs. No Blockage) and $(1 | w)$ represents the random effect.  This method separates the variation caused by the optimization preference (w) from the variation caused by the physics model (Blockage), allowing for a robust estimate of the latter.

\section{Results}

As mentioned above, we used a linear mixed-effects model to assess the effect of the blockage deficit model on the mean AEP. Figure 3 shows the spatial distribution of the optimized turbine positions for $w=0$, $w=0.5$, and $w=1$. The gray circles represent the random initial layout (SmartStart), while the dotted lines trace the displacement vectors to the final optimized positions (crosses). The dashed rectangle indicates the site boundary constraint.

\subsection{Effect on Mean AEP}

The results of the mixed-effects model are presented in Table 1. To interpret these results, we examine the coefficient and the significance level ($p$-value). The fixed effect coefficient $\beta = -0.0624$ represents the average reduction in AEP caused by the blockage model. The $p$-value assesses the statistical significance. It represents the probability of observing a difference of this magnitude purely by random numerical noise if there were no true physical difference. This reduction is not statistically significant ($p = 0.323$), thus we cannot reject the null hypothesis that the observed difference in mean AEP between the two groups is distinguishable from random variability.

\begin{table}[H]
\caption{Results of Mixed-Effects Model for Mean AEP.}
	\centering
	\begin{tabular}{lccc}
	\hline
	Fixed Effects & Estimate &$p$-value \\ \hline
	Intercept ($\beta_0$) & 89.82 & < 0.001 \\
	Blockage Model Effect ($\beta_1$) & -0.0624 & 0.323 \\\hline
	\end{tabular}
\end{table}

\begin{figure}[H]
	\centering
	\includegraphics[width=0.95\textwidth]{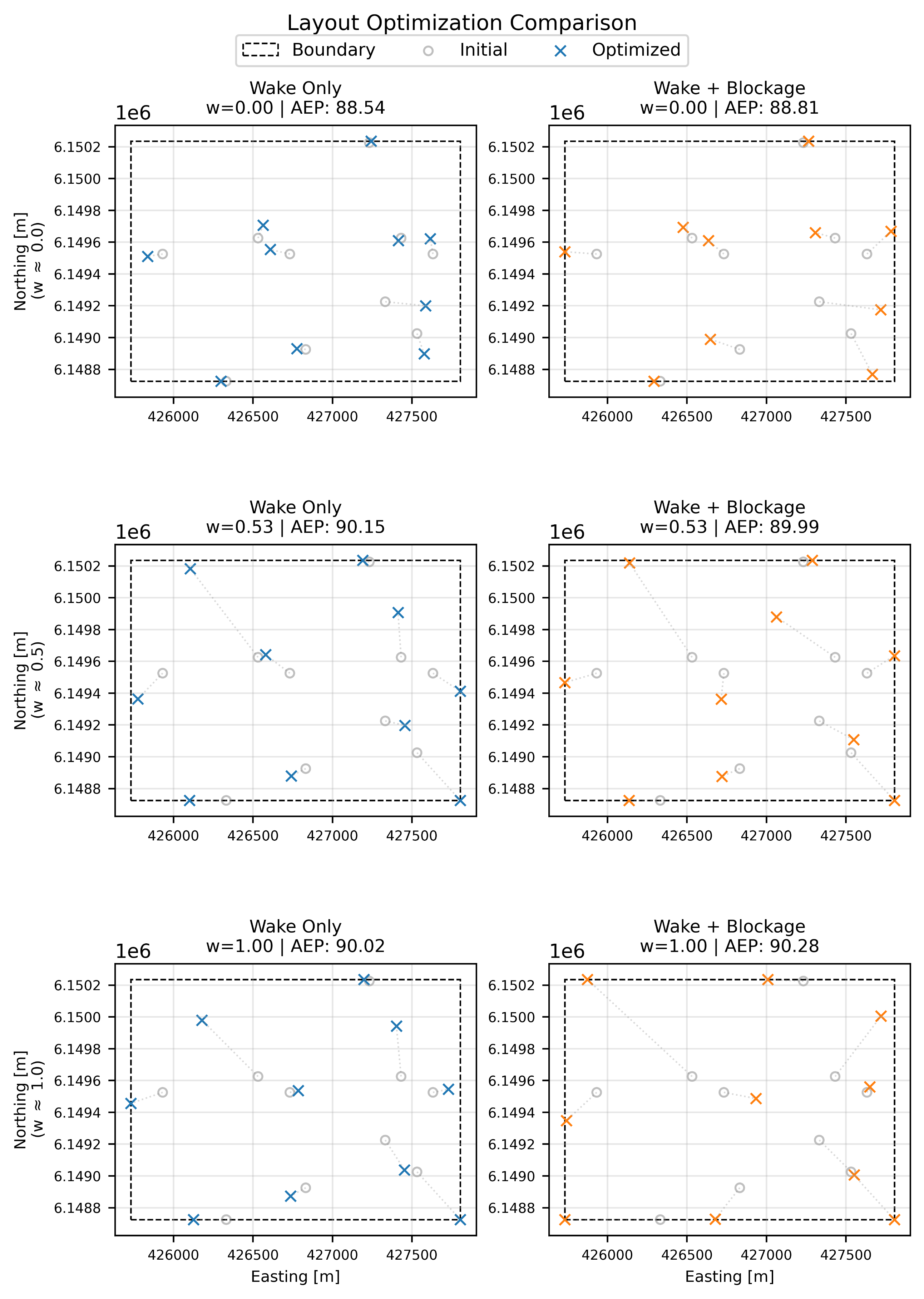}
	\caption{ Comparison of optimized wind farm layouts for the Wake-Only ensemble (left column) and the Wake + Blockage ensemble (right column). Rows correspond to representative objective weights: maximizing mean AEP ($w=0$), a balance trade-off ($w=0.53$), and minimizing AEP uncertainty ($w=1$).}
\end{figure}

Physically, the tendency of the optimized layouts to become coarser (more sparsely distributed) when blockage is included, particularly at $w = 1$, reflects the optimizer’s attempt to decouple the aerodynamic induction zones. Blockage creates an upstream deceleration region. Thus in dense layouts, these individual zones superimpose, amplifying the farm-level blockage penalty and exacerbating the disagreement between wake models. By coarsening the layout, the optimizer mitigates this compounded effect. For actual wind farm design, this presents a critical trade-off: optimizing for robustness against blockage and wake uncertainty dictates larger inter-turbine spacing, which directly conflicts with the industry trend of maximizing turbine density to minimize cable lengths and lease area costs.

Figure 4 provides critical context for this result. Including the blockage model is not a uniform penalty, as it seems to take the role of a stabilizing factor. The layouts optimized with blockage models (Red) appear to show a flatter and more stable AEP profile across the Pareto front than without them (for $0.6 \leq w\leq 0.9$). This suggests that while blockage may not significantly penalize the mean energy production, it may lead to more consistent performance predictions regardless of the specific weighting used.

\begin{figure}[H]
	\centering
	\includegraphics[width=0.95\textwidth]{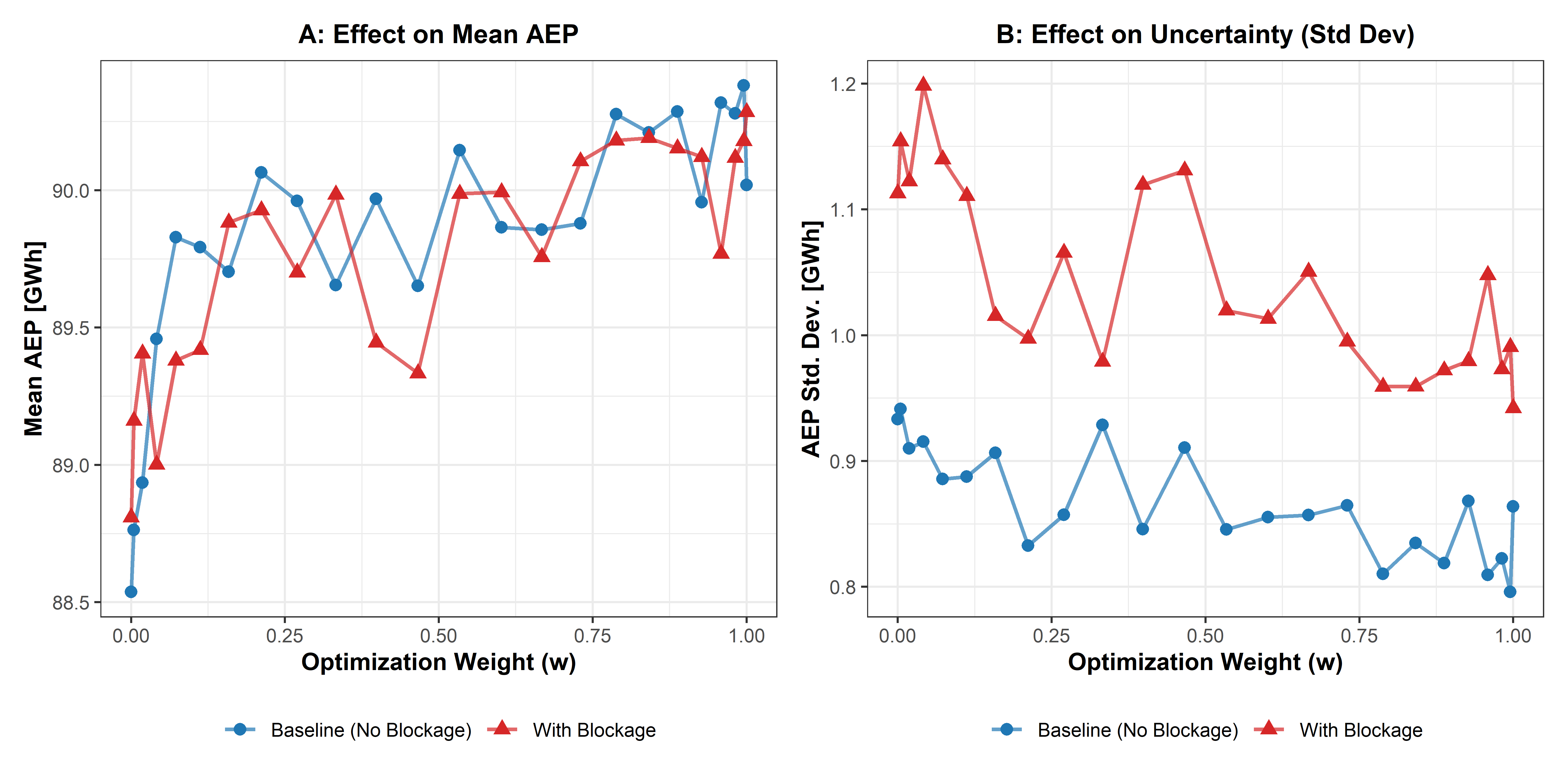}
	\caption{ Effect of blockage models on (A): Mean AEP and (B) Uncertainty (Standard Deviation).}
\end{figure}

\subsection{Effect on Robustness (Uncertainty)}

We applied a similar analysis to the Standard Deviation (SD) of the AEP predictions (Table 2 and Figure 4b). The results show a positive coefficient of $\beta = 0.177$ with high significance $p$-value < 0.001. It means that adding the blockage model increases the disagreement between the engineering wake models by 0.177 GWh.

The marginal $R^2$ value of 0.685 indicates that including blockage physics explained 68.5\% of the variance in uncertainty. While the specific layout configuration ($w$) contributes to the overall variance of 0.85, the inclusion of blockage remains the main contribution of model disagreement, dominating the influence of the optimization weights.

\subsection{Summary}

These statistics quantify a critical trade-off: while blockage models provide a more conservative energy estimate of predicted AEP using Self Similar blockage models (a reduction of 0.0624 GWh, roughly 0.07\%), they introduce significantly higher model-selection uncertainty by increasing the AEP standard deviation by 0.177 GWh. It took more than 11.80 CPU-hours to finish our simulation using blockage deficit models, which is much longer than without them (approximately 1.34 CPU-hours).

\section{Discussion}

\subsection{The Accuracy vs. Certainty Paradox}

Our results highlight a paradox in wind farm modeling. While blockage models are theoretically necessary, specifically the upstream induction effects, to correct bias (possibly reducing the mean AEP estimate), they simultaneously amplify the disagreement (variance) between different wake models. Essentially, our statistical analysis demonstrated that while the inclusion of blockage did not result in a statistically significant reduction in mean AEP ($p > 0.05$), it did result in a highly significant increase in uncertainty ($p < 0.001$). This implies that developers seeking ``more accurate'' models by including blockage must be prepared to accept higher levels of reported uncertainty without necessarily seeing a guaranteed shift in the P50 yield.

\subsection{Financial Impact on Bankability}

Financially, these results suggest a risk to project bankability. The inclusion of blockage physics results in a wider probability distribution (due to the +0.177 GWh uncertainty increase). Even if the central P50 estimate remains stable (as suggested by the non-significant mean shift), the increased variance widens the gap between P50 and P90/P99 estimates. Consequently, P90 and P99 energy yield estimates, which are critical for securing low-interest financing, would be disproportionately lower if blockage is included in the optimization loop.

\subsection{Computational Efficiency and Strategy}

Finally, the computational cost of including blockage in the inner optimization loop is substantial (increasing runtime from $\sim$ 1.34 to $\sim$ 11.8 CPU-hours). Given that the mean AEP correction is relatively minor ($\approx$ 0.07\%), it may be more efficient for developers to exclude blockage during the iterative layout optimization phase and instead apply it as a post-correction step. However, this post-correction approach is insufficient for capturing true global blockage effects, such as flow confinement below a capping inversion layer or wind farm-atmosphere interactions. Capturing those macro-scale phenomena requires the adoption of Multi-Scale Coupled models \cite{Stipa2023, Devesse2024b}. Unlike static post-corrections, these advanced frameworks dynamically resolve the physics between the farm boundary layer and the mesoscale atmosphere, representing a necessary future step for holistic wind farm design.

\subsection{The Impact of Model Compatibility and Calibration}

The significant increase in AEP standard deviation (+0.177 GWh) observed in our results warrants careful physical interpretation. As noted in the GloBE joint statement \cite{RWE2023}, engineering wake models are often implicitly calibrated to account for some blockage effects within their recovery parameters. When an explicit blockage model (like Self Similar) is added without re-calibrating the underlying wake models, it effectively changes the background flow field in ways the wake models were not tuned to handle.

Our results quantitatively reveal a warning of this simple coupling. The variability we observed may not be an inherent property of blockage physics itself, but rather a measure of the incompatibility between the untuned wake models and the blockage field without some careful model tuning and validation with observational SCADA data. Different wake models (e.g., Jensen vs. Gaussian) respond with different sensitivities to the induction gradients introduced by the Self Similar, driving the models further apart.

This highlights the value of the proposed statistical framework. By using the linear mixed model analysis, we were able to isolate and quantify the magnitude of this incompatibility. The framework serves as a diagnostic tool: a statistically significant spike in variance (as seen here) acts as an indicator that the coupled models could be inconsistent and require re-calibration against SCADA before they can be trusted for bankable yield assessments.

\subsection{Limitations and Future Work}

In this study, we restricted the optimization to 10 turbines due to computational constraints. To ensure physical relevance, the site boundary was effectively scaled to maintain density. One limitation is that wake and blockage effects are sensitive to power density (MW/km2). In very dense layouts, the incompatibility effect we observed (wake models disagreeing on the blockage field) may be even more clear. Future work should investigate how this uncertainty scales with wind farm density.

By limiting the blockage model to upstream induction, downstream bypass speed-ups are essentially excluded. Because these speed-ups physically counterbalance a portion of the upstream momentum loss, their omission implies that the absolute magnitude of the wind farm blockage losses in our simulations may be slightly overpredicted. However, the primary trends regarding optimization behavior and wake model variance remain valid. Furthermore, as inter-turbine speed-ups create discontinuities at the wake boundaries, omitting them ensures that the quantified variance strictly isolates the effect of modified inflow deceleration, avoiding confounding numerical artifacts from downstream superposition.

\ack
We would like to thank Jake Badger, Mark Kelly, and Paul van der Laan for interesting discussion and Katayoun Nourbakhsh for proofreading.

\bibliographystyle{unsrtnat}
\bibliography{Torque26_references}

@inproceedings{Mortensen2015,
  author       = {Mortensen, N. G. and Nielsen, M. and J{\o}rgensen, H. E.},
  title        = {Comparison of Resource and Energy Yield Assessment Procedures 2011-2015: What have we learned and what needs to be done?},
  booktitle    = {EWEA Annual Conference and Exhibition 2015},
  year         = {2015},
  organization = {European Wind Energy Association (EWEA)}
}

@article{Oneill2025,
  author    = {O'Neill, N. and R{\'e}thor{\'e}, P. E. and Mouradi, R. S. and Mathieu, A. and Quick, J.},
  title     = {Wind Farm Layout Optimization Accounting for Uncertainty in Model Selection},
  journal   = {Journal of Physics: Conference Series},
  volume    = {3016},
  number    = {1},
  pages     = {012054},
  year      = {2025},
  month     = {May},
  publisher = {IOP Publishing}
}

@article{Branlard2015,
  author  = {Branlard, E. and Gaunaa, M.},
  title   = {Cylindrical vortex wake model: right cylinder},
  journal = {Wind Energy},
  volume  = {18},
  number  = {11},
  pages   = {1973--1987},
  year    = {2015}
}

@inproceedings{Forsting2015,
  author    = {Forsting, A. R. M. and Troldborg, N. and Gaunaa, M.},
  title     = {The flow upstream of a row of aligned wind turbine rotors and its effect on power production},
  booktitle = {Wind Energy 2015},
  year      = {2015}
}

@article{Fisher1919,
  author  = {Fisher, R. A.},
  title   = {The correlation between relatives on the supposition of Mendelian inheritance},
  journal = {Earth and Environmental Science Transactions of the Royal Society of Edinburgh},
  volume  = {52},
  number  = {2},
  pages   = {399--433},
  year    = {1919}
}

@misc{Pedersen2023,
  author = {Pedersen, M. M. and Forsting, A. M. and van der Laan, P. and Riva, R. and Roman, L. A. A. and Risco, J. C. and Friis-M{\o}ller, M. and Quick, J. and Christiansen, J. P. S. and Rodrigues, R. V. and Olsen, B. T. and R{\'e}thor{\'e}, P.-E.},
  title  = {Pywake 2.5.0: An open-source wind farm simulation tool},
  year   = {2023},
  url    = {https://gitlab.windenergy.dtu.dk/TOPFARM/PyWake}
}

@misc{Nygaard2022,
  author = {Nygaard, N. G. and Pedersen, J. G. and Hansen, S. D. and Krastins, P.},
  title  = {Turbopark: A turbulence optimized park model},
  year   = {2022},
  url    = {https://github.com/OrstedRD/TurbOPark/blob/main/TurbOPark\%20description.pdf}
}

@article{Zong2020,
  author  = {Zong, H. and Port{\'e}-Agel, F.},
  title   = {A momentum-conserving wake superposition method for wind farm power prediction},
  journal = {Journal of Fluid Mechanics},
  volume  = {889},
  pages   = {A8},
  year    = {2020}
}

@article{Blondel2020,
  author  = {Blondel, F. and Cathelain, M.},
  title   = {An alternative form of the super-gaussian wind turbine wake model},
  journal = {Wind Energ. Sci.},
  volume  = {5},
  pages   = {1225--1236},
  year    = {2020},
  doi     = {10.5194/wes-5-1225-2020}
}

@article{Bastankhah2014,
  author  = {Bastankhah, M. and Port{\'e}-Agel, F.},
  title   = {A new analytical model for wind-turbine wakes},
  journal = {Renewable Energy},
  volume  = {70},
  pages   = {116--123},
  year    = {2014}
}

@techreport{Jensen1983,
  author      = {Jensen, N. O.},
  title       = {A note on wind generator interaction},
  institution = {Ris{\o} National Laboratory},
  year        = {1983}
}

@article{Bates2015,
  author  = {Bates, D. and M{\"a}chler, M. and Bolker, B. and Walker, S.},
  title   = {Fitting linear mixed-effects models using lme4},
  journal = {Journal of Statistical Software},
  volume  = {67},
  pages   = {1--48},
  year    = {2015}
}

@misc{Rcoreteam2022,
  author       = {{R Core Team}},
  title        = {R: A language and environment for statistical computing},
  organization = {R Foundation for Statistical Computing},
  address      = {Vienna, Austria},
  year         = {2022},
  url          = {https://www.R-project.org}
}

@misc{Riva2023,
  author = {Riva, R. and Liew, J. Y. and Friis-M{\o}ller, M. and Dimitrov, N. K. and Barlas, E. and R{\'e}thor{\'e}, P. and Pedersen, M. M.},
  title  = {Welcome to TOPFARM},
  url    = {https://topfarm.pages.windenergy.dtu.dk/TopFarm2/index.html},
  year    = {2023}
}

@article{Hansen2012,
  author  = {Hansen, K. S. and Barthelmie, R. J. and Jensen, L. E. and Sommer, A.},
  title   = {The impact of turbulence intensity and atmospheric stability on power deficits due to wind turbine wakes at Horns Rev wind farm},
  journal = {Wind Energy},
  volume  = {15},
  number  = {1},
  pages   = {183--196},
  year    = {2012}
}

@misc{RWE2023,
  author = {{RWE} and others},
  title  = {Global Blockage Effect in Offshore Wind (GloBE)},
  year   = {2023},
  url    = {https://www.carbontrust.com/our-work-and-impact/impact-stories/large-scale-rd-projects-offshore-wind/global-blockage-effect-in-offshore-wind-globe}
}

@article{Das1997,
  author  = {Das, I. and Dennis, J. E.},
  title   = {A closer look at drawbacks of minimizing weighted sums of objectives for Pareto set generation in multicriteria optimization problems},
  journal = {Structural Optimization},
  volume  = {14},
  number  = {1},
  pages   = {63--69},
  year    = {1997}
}

@article{Marler2010,
  author  = {Marler, R. T. and Arora, J. S.},
  title   = {The weighted sum method for multi-objective optimization: new insights},
  journal = {Structural and Multidisciplinary Optimization},
  volume  = {41},
  number  = {6},
  pages   = {853--862},
  year    = {2010}
}

@misc{Gribben2019,
  author = {Gribben, B. J. and Hawkes, G. S.},
  title  = {A potential flow model for wind turbine induction and wind farm blockage},
  year   = {2019},
  url    = {https://www.fnc.co.uk/resources/a-potential-flow-model-for-wind-turbine-induction-and-wind-farm-blockage/}
}

@article{Stipa2023,
  author  = {Stipa, S. and Ajay, A. and Allaerts, D. and Brinkerhoff, J.},
  title   = {The Multi-Scale Coupled Model: a New Framework Capturing Wind Farm-Atmosphere Interaction and Global Blockage Effects},
  journal = {Wind Energy Science Discussions},
  pages   = {1--44},
  year    = {2023},
  doi     = {10.5194/wes-2023-75}
}

@article{Devesse2024a,
  author  = {Devesse, K. and Lanzilao, L. and Meyers, J.},
  title   = {A meso--micro atmospheric perturbation model for wind farm blockage},
  journal = {Journal of Fluid Mechanics},
  volume  = {998},
  pages   = {A63},
  year    = {2024},
  doi     = {10.1017/jfm.2024.868}
}

@article{Devesse2024b,
  author  = {Devesse, K. and Stipa, S. and Brinkerhoff, J. and Allaerts, D. and Meyers, J.},
  title   = {Comparing methods for coupling wake models to an atmospheric perturbation model in WAYVE},
  journal = {Journal of Physics: Conference Series},
  volume  = {2767},
  pages   = {092079},
  year    = {2024},
  doi     = {10.1088/1742-6596/2767/9/092079}
}

@article{Branlard2020,
  author  = {Branlard, E. and Quon, E. and Meyer Forsting, A. R. and King, J. and Moriarty, P.},
  title   = {Wind farm blockage effects: comparison of different engineering models},
  journal = {Journal of Physics: Conference Series},
  volume  = {1618},
  pages   = {062036},
  year    = {2020},
  doi     = {10.1088/1742-6596/1618/6/062036}
}

\end{document}